  \documentclass[useAMS,usenatbib]{mn2e}

\usepackage{graphicx}
\usepackage{subfigure}
\usepackage{color}
\usepackage[authoryear]{natbib}
\usepackage{names}

\def\eg{{\it e.g.} }

\begin{document}
\title[]{ The short-lived production of exozodiacal dust in the aftermath of a dynamical instability in planetary systems.}


    \author[A. Bonsor et al.]{Amy Bonsor$^1$\thanks{Email:amy.bonsor@gmail.com}, Sean N. Raymond$^{2, 3}$, Jean-Charles Augereau$^1$ \\
$^1$UJF-Grenoble 1 / CNRS-INSU, Institut de Plan\'{e}tologie et d'Astrophysique de Grenoble (IPAG) UMR 5274, Grenoble, F-38041, France  \\     
$^2$CNRS, UMR 5804, Laboratoire d'Astrophysique de Bordeaux, 2 rue de l'Observatoire, BP 89, F-33271 Floirac Cedex, France\\
$^3$Universit\'{e} de Bordeaux, Observatoire Aquitain des Sciences de l'Univers, 2 rue de l'Observatoire, BP 89, F-33271 Floirac Cedex, France}

   \date{Received ?, 20??; accepted ?, 20??}

\maketitle

\begin{abstract}
Excess emission, associated with warm, dust belts, commonly known as exozodis, has been observed around a third of nearby stars. The high levels of dust required to explain the observations are not generally consistent with steady-state evolution. A common suggestion is that the dust results from the aftermath of a dynamical instability, an event akin to the Solar System's Late Heavy Bombardment. In this work, we use a database of N-body simulations to investigate the aftermath of dynamical instabilities between giant planets in systems with outer planetesimal belts. We find that, whilst there is a significant increase in the mass of material scattered into the inner regions of the planetary system following an instability, this is a short-lived effect. Using the maximum lifetime of this material, we determine that even if every star has a planetary system that goes unstable, there is a very low probability that we observe more than a maximum of 1\% of sun-like stars in the aftermath of an instability, and that the fraction of planetary systems currently in the aftermath of an instability is more likely to be limited to $\le 0.06\%$. This probability increases marginally for younger or higher mass stars. We conclude that the production of warm dust in the aftermath of dynamical instabilities is too short-lived to be the dominant source of the abundantly observed exozodiacal dust.


\end{abstract}

\section{Introduction}
\label{sec:intro}

Observations of dusty, debris discs are common in planetary systems \citep{wyattreview}. $70\mu$m excess, consistent with such a dusty, debris disc, was detected with Spitzer around $\sim15\%$ of sun-like stars, a figure that falls to $\sim4\%$ at $24\mu$m \citep{trilling08}. Although the small dust grains observed in such discs have a short lifetime against either collisions or radiative forces, they are continuously replenished by a collisional cascade that grinds down a reservoir of large parent bodies into small dust grains. Such steady-state collisional evolution can retain cold, outer debris discs over billions of years of evolution. Given that collision rates are higher in denser debris belts, there is a maximum dust level (proportional to the disc luminosity) that can be retained as a function of time, independent of the initial mass of material in the parent reservoir \citep{Wyatt07hot}. Although the exact value of this maximum is not well constrained \citep{Wyatt07hot, lohne, Gaspar2008}, most cold, outer debris discs are significantly fainter than this maximum \citep{wyattreview}, however, there are many observations of warm or hot, inner dust belts that exceed this maximum level \citep{Wyatt07hot, etacorvi}. Such warm dust belts are commonly known as exozodi, referring to their similarity with the Solar System's zodiacal cloud. Although it should be noted that in order to be detectable, an exozodi must have a fractional luminosity several orders of magnitudes higher than our Solar System's zodiacal cloud \citep{Absil06, Absil08, Millan-Gabet2011}.

Recent observations suggest that exozodiacal dust may be commonplace. There are a few detections of systems with extremely high levels of exozodiacal dust, where it has been possible to obtain spectra with Spitzer (\eg $\eta$ Corvi \citep{resolveHD69830, Lisse12}, HD 172555 \citep{Lisse2009}). \cite{Lawler2011} find only $\sim1\%$ of their sample have excess emission at 8.5-12$\mu$m with IRS. They were, however, limited to detections brighter than 1000 times the level of zodiacal dust in our Solar System. Similarly, using WISE \cite{Kennedy_Wise} find that fewer than 1/1000 stars have $12\mu$m excess more than a factor of 5 above the stellar photosphere. Near and mid-infrared interferometry is better suited to probing the inner regions of planetary systems.  \cite{Millan-Gabet2011} find a mean level of exozodiacal dust less than 150 times the level of that in our Solar System for their sample of 23 low dust stars in the N-band, whilst detecting exozodiacal dust in two stars already known to host dusty material ($\eta$ Corvi and $\gamma$ Oph). A survey at shorter wavelengths (K-band 2-2.4$\mu$m) using CHARA/FLUOR \citep{Absil2013} found K-band excess for $28^{+8}_{-6}\%$ of their sample of 40 single stars, chosen to be spread equally in spectral type and age and half of which also have cold excess emission. \cite{Absil2013} found no correlation with age, but a significant difference between their sample of A stars, $50^{+13}_{-13}\%$ of which exhibit K-band excess, compared to only $18^{+9}_{-5}\%$ of their sample of FGK stars.

Detailed modelling of individual exozodis finds that the emission is best reproduced by populations of very small grains, in many cases smaller than the limiting diameter at which grains are blown out of the system by radiative forces \citep[e.g.][]{lebreton2013,Defrere_Vega}. This implies a very short lifetime for the observed dust and even if it were to be replenished by collisions between larger parent bodies, the lifetime of the parent bodies remains relatively short, compared to the age of the systems. There has been much discussion in the literature as to their origin, in particular, whether or not such dust discs are a transient or steady-state phenomenon \citep[e.g][]{Wyatt07hot, bonsor_exozodi,etacorvi, Fujiwara2009, Moor2009, Olofsson2012}. Detailed analysis of the spectral composition of a couple of targets provides evidence that we are witnessing the recent aftermath of a high-velocity collision \citep{Lisse12,Lisse2009}. This could be a rare event, such as the Earth-Moon forming collision \citep{Jackson2012} or an equivalent to the Solar System's Late Heavy Bombardment (LHB) \citep[e.g.][]{Lisse12,Wyatt07hot}. \cite{Booth09} use predictions for the levels of dust production following our Solar System's LHB, compared with Spitzer observations at $24\mu$m to constrain the maximum fraction of sun-like stars that undergo a LHB to be $<12$\%. Alternatively it has been suggested that the observed dusty material could be linked in a more steady-state manner with an outer planetary system \citep{Absil06,resolveHD69830, bonsor_exozodi} or even that we are observing the increased collision rates of a highly eccentric distribution of planetesimals at pericentre \citep{ecc_ring}.

In this work we assess the possibility that the stars with exozodiacal dust are planetary systems that have been observed in the aftermath of a dynamical instability. Dynamical instabilities trigger a phase of planet-planet gravitational scattering that usually culminates with the ejection of one or more planets, or at least a large-scale change in the planetary system architecture.  The planet-planet scattering model can naturally explain several features of the observed giant exoplanets, notably their broad eccentricity distribution \citep{Chatterjee08, Juric2008, Raymond2010} and their clustering near the Hill stability limit \citep{Raymond09}. If this model is correct, the planets we observe represent the survivors of  instabilities, which occur in at least half and up to $>$90\% of all giant planet systems.  The Solar System's LHB is a very weak example of such an instability, as it is thought that gravitational scattering did occur between an ice giant and Jupiter, but not between gas giants \citep{Morbidelli2010}.

Dynamical instabilities among giant planets have drastic consequences for the planetesimal populations in those systems.  As the planets are scattered onto eccentric orbits they perturb small bodies both interior and exterior to their starting locations \citep{Raymond2011, Raymond2012}. Planetesimals' eccentricities are quickly and strongly excited.  On their way to ejection, many planetesimals spend time on eccentric orbits with close-in pericentre distances, where they have the potential to produce the emission observed in systems with exozodi. However, the increased levels of dust, particularly in the inner regions, will only be sustained for a short time period following the instability, limited by the collisional and dynamical lifetime of the material. We call this time period for which increased levels of dust are sustained in the inner system, $t_{zodi}$.

If all exozodis were to be planetary systems observed in the aftermath of an instability, we can estimate the fraction of stars where exozodiacal dust is observed, $Y$, by considering the fraction of systems that we are observing within $t_{zodi}$ of an instability. We make the naive assumption that there is an equal probability for an instability to occur at any point during the star's main-sequence lifetime, $t_{MS}$ and similarly, that there is an equal probability to observe a star at any point during $t_{MS}$. This leads us to conclude that the probability of observing systems within $t_{zodi}$ of an instability is given by the ratio $\frac{t_{zodi}}{t_{MS}}$. More generally, the distribution of instabilities as a function of time would need to be considered, alongside the distribution of values of $t_{zodi}$ and $t_{MS}$. For completeness we also include two factors that we generally assume to be close to one, the first, $f_{pl}$, indicates the fraction of stars that have planetary systems, with both giant planets and a planetesimal belt and, the second, $N_{inst}$, the number of `exozodi-producing' instabilities each planetary system may have. Thus:

\begin{equation} 
Y =f_{pl} \; N_{inst}  \; f_{unstable} \;  \frac{t_{zodi}  }{t_{MS}}, 
\label{eq:y}
\end{equation}
where any dependence on the architecture of the planetary system has been integrated over, $f_{unstable}$ refers to the fraction of planetary system that suffer an instability, and we envisage that all the parameters are a function of the spectral type of the star.

In this work we use N-body simulations to determine the lifetime of high dust levels following an instability, $t_{zodi}$, and in doing so use Eq.~\ref{eq:y} to illustrate that it is not possible for all, if any, of the systems where exozodiacal dust is detected to be in the aftermath of an instability. We start by describing our simulations in \S\ref{sec:sim}, which we follow by a discussion of the manner in which we calculate the parameter, $t_{zodi}$, illustrated by an example simulation in \S\ref{sec:scatt}. In \S\ref{sec:tzero} we discuss the range of values of $t_{zodi}$ obtained in our simulations and in \S\ref{sec:dyn} the implications of these results regarding the fraction of systems that we are likely to be observing in the aftermath of an instability. This is followed by a discussion of the approximations used in our calculations, as well as the consequences of changing various parameters in \S\ref{sec:discussion}. We finish with our conclusions in \S\ref{sec:conclusion}.





\section{The simulations}
\label{sec:sim}
We make use of a large database of simulations of planet-planet scattering in the presence of planetesimal discs, used for previous work on the planet-planet scattering mechanism and its consequences \citep{Raymond2008, Raymond09, Raymond2009b, Raymond2010}. These simulations provide many examples of dynamical instabilities and the consequences of such instabilities on small bodies in the planetary system. Whether or not these simulations are representative of the population of real planetary systems, they provide a good range of realistic examples of planetary systems in which instabilities occur, from which it is possible to better understand the effects of dynamical instabilities on planetary systems. In this work, we focus particularly on the scattering of small bodies into the inner regions of the planetary systems, in a manner that could potentially produce the observed warm dust. These simulations are for solar-type stars, with outer belts at $\sim10$AU, although we will discuss further in \S\ref{sec:discussion} some simple scaling relations that can be used to estimate the behaviour in other systems.

The simulations in the database contain three giant planets interior to a planetesimal disc. The planetesimal disc extends from 10 to 20AU and contains $50M_\oplus$, divided equally among 1000 planetesimals. The planetesimals are considered to be `small bodies', thus, their gravitational interactions with the giant planets, but not  each other, are taken into account. The planetesimals followed an $r^{-1}$ surface density profile and were given initial eccentricities randomly chosen in the range $0-0.02$ and inclinations between $0-1^\circ$. The outermost giant planet was placed 2 Hill radii interior to the inner edge of the disc. Two additional giant planets extended inward, separated
by $\Delta$ mutual Hill radii $R_{H,m}$, where $\Delta = 4 - 5$ and :
\begin{equation}
R_{H,m} = \left(\frac{a_1 + a_2}{2}\right)\left(\frac{M_1 + M_2}{3M_*}\right)^{1/3},
\end{equation}
 where, $a$ refers to the planets’ semi-major axes and $M$ to their masses, subscripts 1 and 2 to the individual planets, and $M_*$ to the stellar mass.

The planets were given initially circular orbits with inclinations
randomly chosen between zero and $1^\circ$. Each simulation
was integrated for 100 Myr using the Mercury hybrid symplectic
integrator \citep{chambers99} with a timestep of 5-20
days, depending on the simulation. The positions and velocities of all particles were recorded every $10^5$yrs, in addition to the time at which a close encounter occurred (to the nearest timestep). 
 Collisions were treated as inelastic mergers conserving
linear momentum. Particles were considered to be ejected if
they strayed more than 100 AU from the star. 

In our simulations planet masses were randomly selected according to the distribution of planet masses observed \citep{Butler2006} :
\begin{equation}
\frac{dN}{dM} \propto M^{-1.1}
\end{equation}
We consider two sets of simulations, in the first, that we call {\tt low mass}, planet masses were selected between $10M_\oplus$ to $3M_J$, whilst in the second, {\tt high mass} planet masses were selected between $1M_{sat}$ and $3M_J$ (see \cite{Raymond2010} for more details)\footnote{In \cite{Raymond2010} the simulations {\tt low mass} and {\tt high mass} were referred to as {\tt mixed2} and{ \tt mixed1}.  }. Both sets of simulations contain 1,000 individual simulations, although as we discuss later only 223 and 201 were suitable to use in this work. These simulations have previously been used to reproduce the observed eccentricity distribution of extra-solar planets, which was found to be better reproduced by the {\tt high mass} simulations \citep{Raymond2010}.



\begin{figure}
\includegraphics[width=0.48\textwidth]{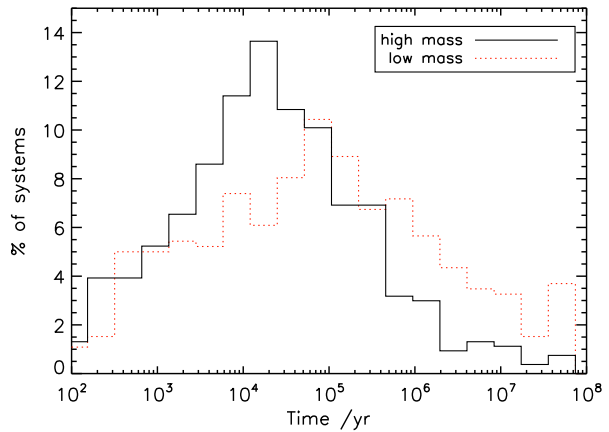}

\caption{A histogram to show the times at which instabilities occurred in our simulations. The difference in initial conditions between {\tt high mass} and {\tt low mass} is explained in \S\ref{sec:sim}. It should be noted that these simulations were designed such that instabilities occur early in the system's evolution. } 
\label{fig:hist_tinst}

\end{figure}

\begin{figure}
\includegraphics[width=0.48\textwidth]{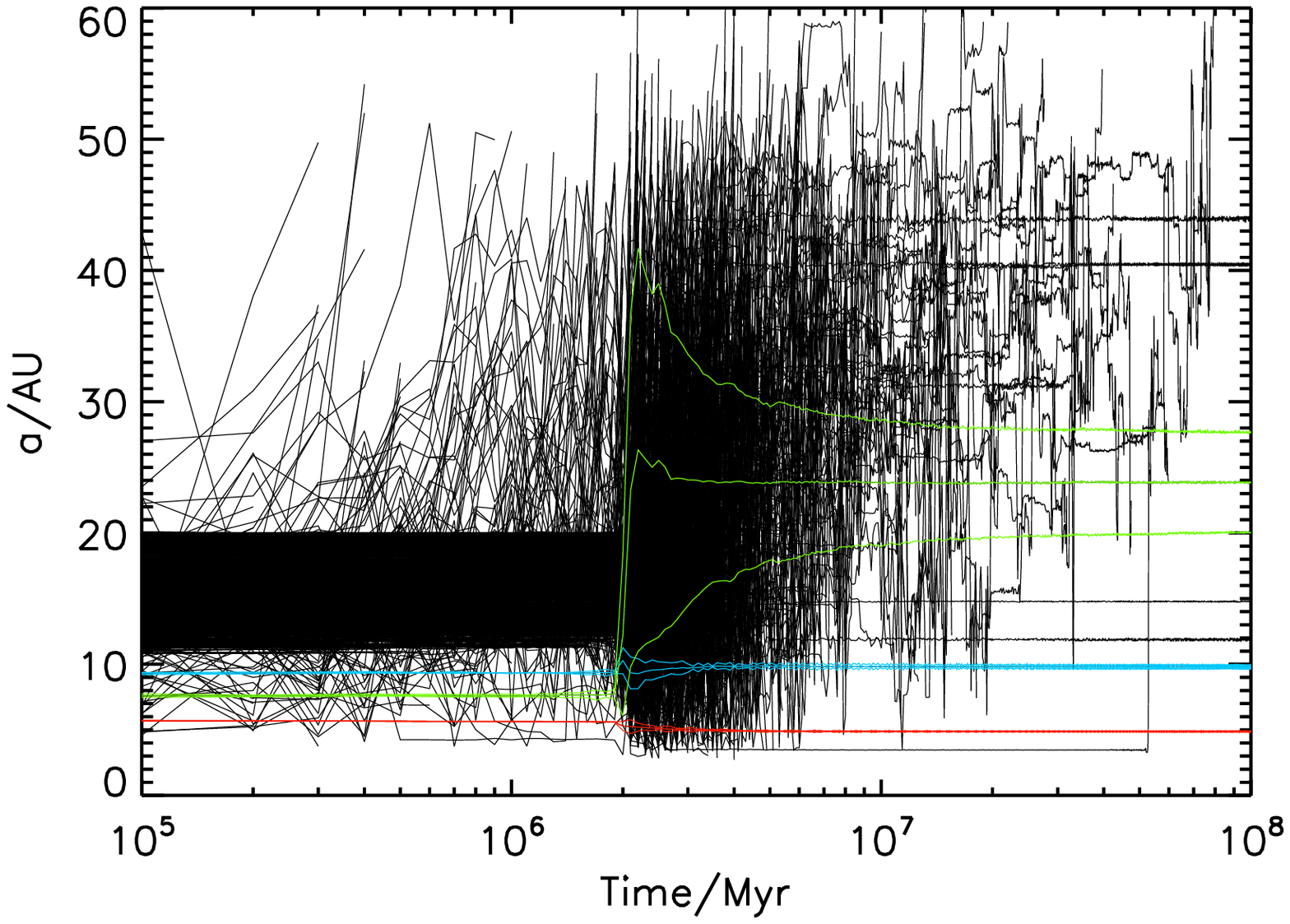}
\includegraphics[width=0.48\textwidth]{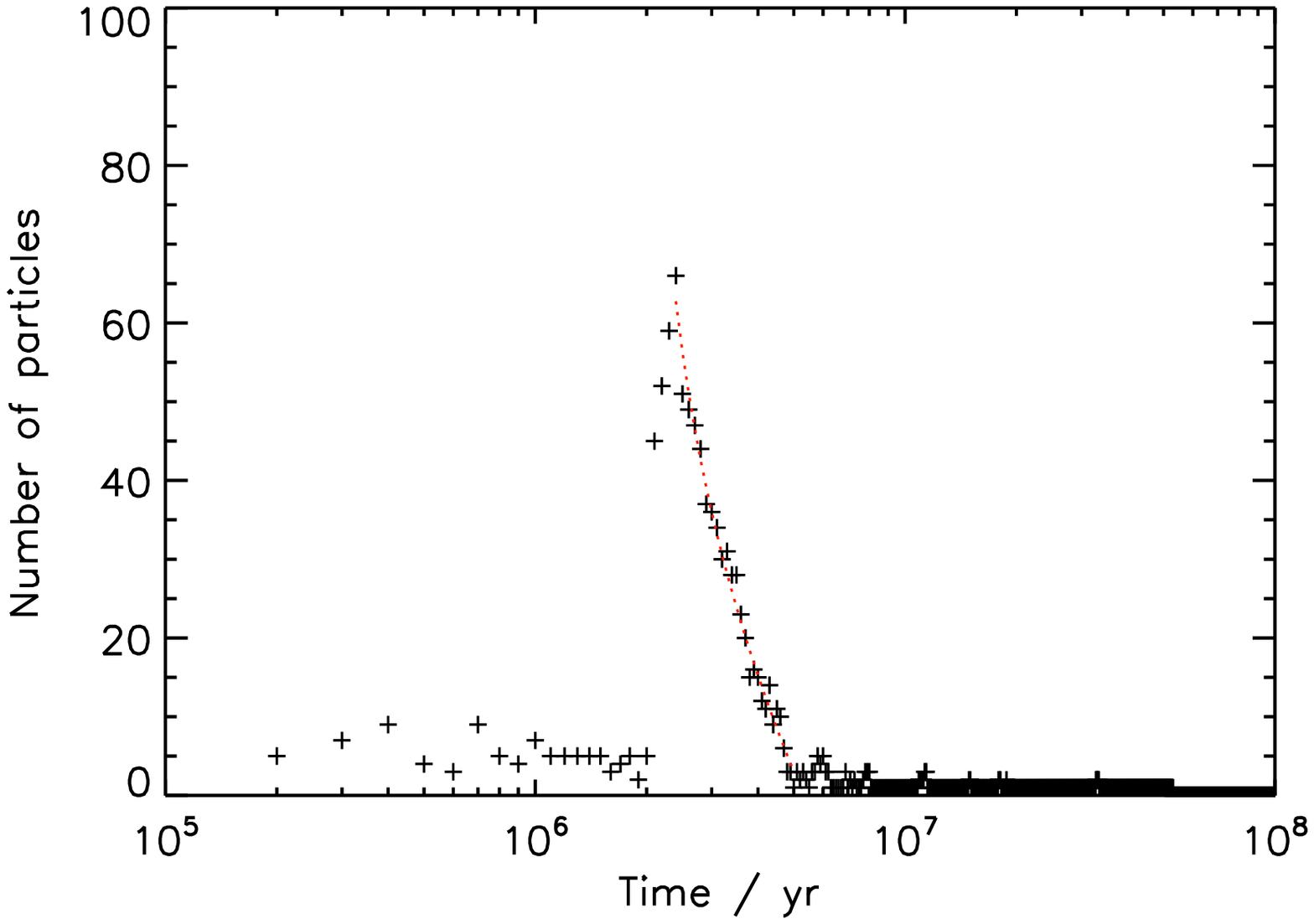}
\caption{ The results of an example simulation with 58.3, 23.5 and 166$M_\oplus$ planets, that started at 9.3, 7.5 and 5.7AU, respectively. The top panel shows the variation of the semi-major axis of all the planets (in red, blue and green) and planetesimals (black) with time. The pericentre and apocentre of the planets are also plotted. The bottom panel shows the number of particles found inside of $r_{lim}=4.7$AU as a function of time, counted at snapshots separated by $10^5$yrs. The dotted line is the fit to the data calculated using Eq.~\ref{eq:ntot}} 
\label{fig:533}
\end{figure}

\section{Do dynamical instabilities produce the observed population of exozodis?}

\subsection{The lifetime of the aftermath of an instability, \lowercase{$t_{zodi}$}}

\label{sec:scatt}
In the database we find a wide range of different instabilities, some occurring very early in the system's evolution, others after millions of years of evolution, some merely leading to changes in the planet's orbital parameters, whilst others to the ejection of one or more planets, {\it etc}. We define an instability to be a close encounter between any pair of planets in our simulations. Fig.~\ref{fig:hist_tinst} shows a histogram of the times after the start of the simulation at which the instability occurred.

In this work we are not concerned by the details of the different instabilities, rather we focus on the aftermath of the instability. In almost all cases, destabilising the planet's orbits means an increase in the scattering of smaller bodies. Many of the scattered bodies are ejected from the planetary system, but some are scattered into the inner regions of the planetary system. Thermal emission from small dust grains produced by the collisional evolution of this scattered material may be detectable as an `exozodi'.


In this work, we consider particles scattered into the inner regions of the planetary system, as a proxy for an exozodi. The best way to illustrate this is to consider an example system. In the top panel of Fig.~\ref{fig:533} we show a simulation in which three planets of masses 58.3, 23.5 and 166$M_\oplus$, started at 9.3, 7.5 and 5.7AU, respectively. After 2Myr, the middle planet is kicked into the outer belt, where it proceeds to clear the belt of material by scattering all planetesimals that approach it. We track the number of particles that are scattered into the inner regions of the planetary system as a function of time. The inner regions are defined by considering particles that are scattered interior to all three planets, using a limiting radius, $r_{lim}$, given by the minimum pericentre of any of the planets throughout the 100Myr evolution considered by our simulations. The number of particles, $N_{in}$, found interior to this radius is shown in the bottom panel of Fig.~\ref{fig:533} as a function of time. We count the number of particles with $q<r_{lim}$ at each timestep for which information regarding the system is stored, every $10^5$yrs. There is a significant increase in $N_{in}$ following the instability, after which $N_{in}$ falls off rapidly with time, as the planet clears the outer belt. We fit a simple straight line dependence to this decrease in the number of particles in this region (in log-time space):

\begin{equation}
N_{in}= A+\alpha\log(\frac{t - t_{inst}}{yr}) \; \; \; \; \mathrm {for } \;\;\; t>t_{inst},
\label{eq:ntot}
\end{equation}
where $t$ is the time since the start of the simulation, $t_{inst}$ the time since the start of the simulation at which the instability started, defined by when the planets first had a close encounter and, $A$ and $\alpha$ coefficients found from the fit to the simulations. For this simulation, we find $A=459$ and $\alpha=-70$. 
In order to obtain a good fit, we include only data points between the maximum in the number of particles in the inner regions and when the number first falls below 3 particles.

We can now calculate the maximum possible lifetime of an exozodi following the instability, $t_{zodi}$, by assuming that this is given by the maximum time period for which the mass of material (number of particles) interior to $r_{lim}$ is non-zero. Using our straight-line fit, Eq.~\ref{eq:ntot}, 
\begin{equation}
t_{zodi}= 10^{-A/\alpha} \; \; yr.
\label{eq:zero}
\end{equation}

For this example simulation we find a maximum possible lifetime for the exozodi of $t_{zodi}=3.4$Myr. This is the maximum possible dynamical lifetime of the scattered material. This ignores the conversion of the scattered material into the dust population that is detected and the observational constraints on detecting this dust population, both of which are discussed in further detail in \S\ref{sec:discussion}.

We note here, that the resolution of our results is limited by the fact that output information was only stored every $10^5$yrs by our simulations. Tests for a sub-sample of simulations, using information stored for the orbital parameters before and after close encounters, found that the time at which a scattered particle first enters the inner regions of the planetary system may have been overestimated by as much as 30-50\%, although for most simulations the over-estimation was only on the level of $1-2\%$. This means that it is possible that we have marginally overestimated the maximum possible lifetime of an exozodi following an instability, $t_{zodi}$. Overestimation of the maximum only strengthens our argument that this is very much a maximum possible lifetime for an exozodi and that the real lifetime is likely to be significantly shorter. 

Another assumption made in our simulations is that the lifetime of particles scattered interior to 0.5AU is always less than $10^5$yrs. This assumption should be relatively robust, given that the collisional lifetime of material so close to the star is short, even if the dynamical lifetime may be longer. The time-step used in our simulations is not sufficiently small to resolve interior to 0.5AU.


\begin{figure}
\includegraphics[width=0.48\textwidth]{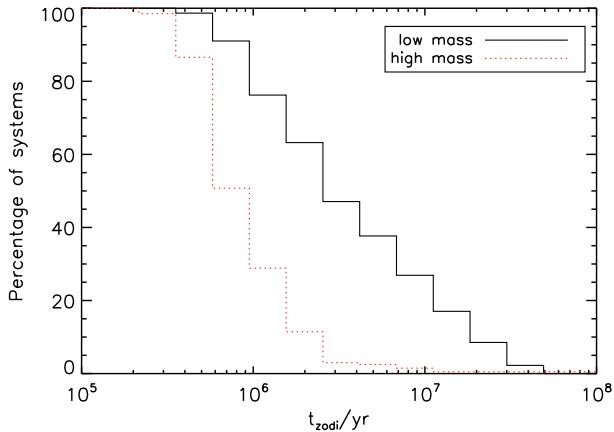}

\caption{The percentage of systems for which material continued to be scattered inwards on timescales greater than the plotted value, or a (reversed) cumulative histogram showing the fraction of systems for which $t_{zodi}$  (calculated using Eq.~\ref{eq:ntot} and Eq.~\ref{eq:zero}) is greater than the plotted value. This falls off sharply with time. There are 223 and 201 simulations, in {\tt low mass} and {\tt high mass}, respectively, once all simulations in which planetesimal driven migration have been discarded in order to isolate the effects of instabilities. The maximum $t_{zodi}$ for the {\tt high mass} ({\tt low mass}) simulations was 11Myr (47Myr) and the median value 0.7Myr (2.8Myr). } 
\label{fig:zero}
\end{figure}

\subsection{Determining \lowercase{$t_{zodi}$} in our simulations}
\label{sec:tzero}
Using the method described in the preceding section, we determine the value of $t_{zodi}$ for every simulation in our database. Any simulations in which there is significant migration of the planets following an instability are neglected. This is defined by a change in semi-major axis of the outer planet of larger than a factor 1.1 between $10^5$years after the instability and the end of the simulations. Planetesimal driven migration will be investigated in a separate article (Bonsor \& Raymond, in prep) and here we prefer to consider the pure affects of dynamical instabilities. We also neglect simulations in which too few particles were scattered into the inner regions in order to fit a straight line to the number of particles found interior to $r_{lim}$ as a function of time. This leaves us with a sample of 223 and 201 simulations, in {\tt low mass} and {\tt high mass}, respectively.

A histogram of $t_{zodi}$ is shown in Fig.~\ref{fig:zero}. The first clear conclusion to be made from this figure is that there is a steep fall-off in the number of systems with larger values of $t_{zodi}$. This plot shows that, whilst for almost every simulation there is an increase in material in the inner regions of a planetary system following an instability, the flux of particles into the inner regions falls off steeply with time after the instability. In fact, this plot shows that for none of our simulations was the value of $t_{zodi}$ larger than 50Myr (11Myr) for the {\tt high mass} ({\tt low mass}) simulations and in the majority of cases significantly less, with a median value of 2.8Myr for the {\tt low mass} and 0.7Myr for the {\tt high mass} simulations.

\begin{figure}
\includegraphics[width=0.48\textwidth]{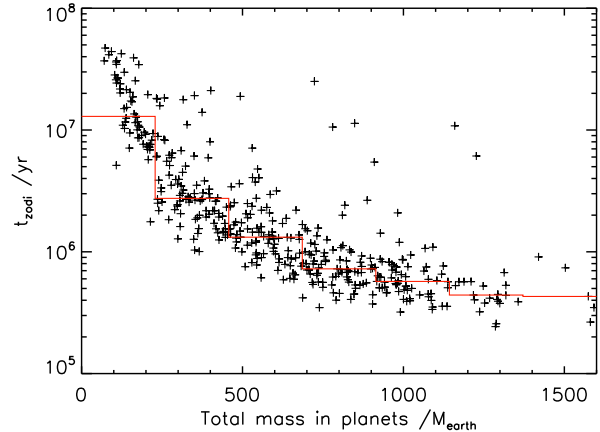}

\caption{The decrease in the timescale on which material survives following an instability ($t_{zodi}$) as a function of the total mass of the three planets in each simulation. The dots refer to individual simulations, whilst the red line indicates the median $t_{zodi}$. There is a clear decrease in $t_{zodi}$ with increasing planet mass.} 
\label{fig:masspl_zero}

\end{figure}

$t_{zodi}$ is inversely proportional to planet mass, as shown in Fig.~\ref{fig:masspl_zero}. High mass planets rapidly clear material in their vicinity, whilst lower mass planets can take significantly longer timescales before all nearby particles are scattered. This explains the difference between the {\tt high mass} and {\tt low mass} simulations in Fig.~\ref{fig:zero}. The spread in this plot results from the fact that the total mass of planets does not always correlate with the mass of the planet that is responsible for the scattering.


If an instability occurs early in our simulations, before the giant planets have had sufficient time to clear their chaotic zones of material, there is additional material that can potentially be scattered into the inner regions of the planetary system. This means that systems with early instabilities have a higher probability of producing a brighter exozodi. Analysis of our simulations found this changes the mass of material scattered into the inner regions, but that $t_{zodi}$ is largely unaffected. 

\subsection{Fraction of stars with observable exozodi?}
\label{sec:dyn}
As discussed in the Introduction, the aim of this work is to determine the probability of observing a planetary system in the aftermath of an instability and thus, to determine whether or not the population of stars observed with exozodiacal dust are likely to be in the aftermath of an instability. Now that we have determined the value of $t_{zodi}$, we can use Eq.~\ref{eq:y} to assess this. Firstly, we consider the absolute maximum value that could possibly be obtained for $Y$ and show that even this maximum is significantly below the observed frequency of exozodi. Then, we use more reasonable assumptions to approximate the probability of detecting a system in the aftermath of instability.

To calculate the maximum possible value of $Y$ we assume that every star has a planetary system, with a planetesimal belt and three giant planets ($f_{pl}=1$), and that every such planetary system has an instability, and that this instability has an equal probability to occur at any point during the star's main-sequence lifetime ($f_{unstable}=1$). We then assume that every planetary system can only have one instability that could produce a detectable exozodi ($N_{inst}=1$), both because multiple instabilities are improbable and in most cases, the first instability would have cleared the planetary system of material previously. This assumption will be discussed in further detail in \S\ref{sec:oneinstability}. We then consider the maximum possible value of $t_{zodi}$ found in any of our simulations, which was 50Myr and a minimum main-sequence lifetime of 5Gyr, for sun-like stars  \citep{sse}. 
This leads to a maximum possible value of $Y$ of $1\%$. Even such a maximum value is clearly significantly lower than the $18\%$ of FGK stars detected by \cite{Absil2013} and given the maximal nature of the assumptions made during its calculation, would even struggle to explain the 1\% detection rate of \cite{Lawler2011}.

If we then take the median value of $t_{zodi}$ from our simulations, we can make a better approximation of the fraction of systems that we might expect to observe in the aftermath of an instability, $Y$. We retain the assumption that every star has a planetary system that undergoes one instability, with equal probability to occur during its main-sequence lifetime of 5Gyr. For the {\tt low mass} simulations with their median value of $t_{zodi}$ of 2.8Myr, we find $Y=0.06\%$, and for the {\tt high mass} simulations, that better fit the observed eccentricity distribution of extra-solar planets, with their median value of $t_{zodi}$ of 0.7Myr, we find $Y=0.01\%$. In other words, this implies that we would need to survey thousands of stars in order to stand a reasonable chance of detecting a single system with exozodiacal dust in the aftermath of a dynamical instability.




\section{Discussion}
\label{sec:discussion}
We have presented arguments that lead us to conclude that the majority of exozodis are not planetary systems observed in the aftermath of a dynamical instability. This conclusion is based on our calculation of the maximum possible fraction of stars that could be detected in the aftermath of an instability, $Y$. We determined that the lifetime of scattered material following an instability, $t_{zodi}$, is short, and therefore, $Y$ is significantly less than the fraction of systems detected with exozodiacal dust. We assumed that every planetary system can have a maximum of one exozodi-producing dynamical instability ($N_{instability}=1$) and our simulations considered sun-like stars ($1M_\odot$), with 3 planets and a planetesimal belt between 10 and 20AU. We now discuss the validity of our conclusions. 

\subsection{Fate of scattered material}

 The most fundamental consideration missing from our calculation regards the collisional evolution of the scattered material and the manner in which this leads to the population of small dust that could be detected as an exozodi. Such a calculation is very complicated and current state of the art codes are just approaching the point where they can start to couple dynamical and collisional simulations \citep{colgrooming09, Thebault_binary2012,Kral2013}. Fortunately the precise manner in which the dusty material is produced and evolves does not affect our results, as collisions are likely to destroy the scattered particles on timescales significantly shorter than their dynamical lifetime. Thus, the real value of $t_{zodi}$ is likely to be significantly shorter than those quoted here, adding weight to our conclusion.

Another possibility that has been suggested in the literature is that the observed exozodi refer to nano grains trapped in the magnetic field of the star \citep{Su2013,Czechowski2010}. We acknowledge here that such a possibility could increase the lifetime of an exozodi significantly beyond the values of $t_{zodi}$ calculated from our simulations, but that such a scenario requires further detailed investigation.


\subsection{How representative are our simulations?}

Our conclusions depend on whether or not our median and maximum values for $t_{zodi}$ are representative of the real population of planetary systems. In our favour is the fact that our simulations are able to reproduce many of the observed features of the exoplanet distribution, including the mass distribution, the eccentricity distribution and to some extent the orbital distribution, although our giant planets are more distant than those that have been currently observed using the RV technique \citep{Raymond2010}. We have a reasonably good statistical sample (201 + 223 simulations with instabilities) and we see a wide range of different {\it types} of instabilities, from relatively gentle close encounters to dramatic scattering events that eject one or more planets.

We are, however, limited in the number of planets that we consider, as well as the initial configurations of the planets and discs. Planetary systems with more than 3 planets could potentially suffer longer lasting instabilities that involve more than 2 planets. The separation of the planets by $4-5R_H$ is consistent with planet formation models in which planets become trapped temporarily in MMRs during the gas disc phase \citep{Snellgrove2001, Lee2002, Thommes2008}. If planetesimal belts exist at larger radii than our choice of 10-20AU, then $t_{zodi}$ could be increased. \cite{trilling08} only find debris discs orbiting sun-like stars with radii less than $\sim30$AU. Given that scattering timescales scale approximately with orbital periods, the maximum increase in $t_{zodi}$ that this could lead to is given by approximately a factor of $\left(\frac{30}{10}\right)^{3/2}\sim 5.2$, which leads to a median frequency of 0.3\% ({\tt low mass}) or 0.05\% ({\tt high mass}). Similarly the limited width of our belt, could artificially retain $t_{zodi}$ at low values. Test simulations found that increasing the belt width to 20AU, using the {\tt wide disk} simulations from \cite{Raymond2012}\footnote{As these simulations were designed to investigate terrestrial planet formation, they also contained $9M_\oplus$ in 50 planetary embryos and 500 planetesimals, distributed from 0.5 to 4AU, according to a radial surface density profile $\Sigma \propto r^{-1}$. These should not affect the broad results of our simulations.} resulted in a median lifetime, $t_{zodi}$, of 4.7Myr, consistent with our {\tt low mass} simulations.

\subsection{How valid are our assumptions regarding instabilities?}

\begin{figure}
\includegraphics[width=0.48\textwidth]{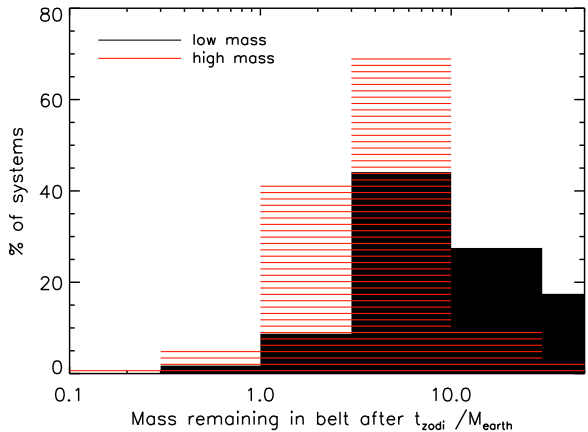}

\caption{A histogram to show the mass that remains in the belt a time $t_{zodi}$ after an instability for our simulations. We define the belt as particles found exterior to 10AU on stable orbits with $e<0.99$. More than $10M_\oplus$ (1/5 of the original belt) of material remained in the belt in less than 45\% (12\%) of the {\tt low mass} ({\tt high mass}) simulations. This ignores the collisional evolution of the belt and is dependant on our initial belt mass of $50M_\oplus$ and location at 10-20AU. } 
\label{fig:hist_remain}

\end{figure}

Firstly, we assume that every planetary system can only have {\it one} instability that leads to a detectable exozodi. Even if some planetary systems were to have 2 or 3 instabilities that produced exozodis this would only increase $Y$ by a factor of 2 or 3 (see Eq.~\ref{eq:y}), and our conclusion would remain unaltered. We found no simulations in which more than two instabilities occurred\footnote{We define a second instability to be a close encounter(s) between planets later than a time $t_{zodi}$ following the first close encounter.}. In addition to this, generally insufficient material remained in the outer belt following the first instability, such that the second instability could scatter sufficient material to be detectable. Fig.~\ref{fig:hist_remain} shows that in less than 45\% (12\%) of our {\tt low mass} ({\tt high mass}) simulations, more than $10M_\oplus$ of material survived in the outer belt, a time $t_{zodi}$ after the instability.

Secondly, we assumed that instabilities are linearly distributed in time, when in reality they are likely to be biased to younger ages (as was shown in Fig.~\ref{fig:hist_tinst}). This could increase the detection rate for younger stars following instabilities, but does not help explain the large quantities of dust observed in old (Gyr) planetary systems such as $\tau$ Ceti  \citep{diFolco07}. 

Thirdly, we assume that every star has a planetary system ($f_{pl}=1$) that has an instability ($f_{unstable}=1$). These are clearly maximal assumptions and whatever the real statistics, they allow us to clearly rule out the ability of instabilities to produce exozodi. Observations are currently not sensitive enough to determine whether or not all stars have planetary systems, but at the very least 5-20\% \citep{Marcy2008,Howard2010, Bowler2010, Mayor2011}, potentially up to 62\%, considering micro-lensing observations that are sensitive to super-Earths \citep{Cassan2012}, of stars possess planets and 10-33\% of stars have debris discs \citep{trilling08, Su06}. As regards whether all of these planetary systems suffer instabilities, the only planetary system where we have any information is our own Solar System where we think an instability triggered the Late Heavy Bombardment \citep{Gomes2005}. The only real clues that we have for exo-planetary systems come from the broad eccentricity distribution observed. To reproduce this distribution using simulations of planet-planet scattering requires that roughly 75-95\% of giant planet systems go unstable \citep{Chatterjee08, Juric2008, Raymond2010}. Nonetheless, a modest fraction of giant planet systems (up to roughly 25\%) may remain stable for long timescales.



\subsection{Observable signatures of instabilities}

Another critical argument in support of a short lifetime for the dusty aftermath of dynamical instabilities regards our limited ability to detect dust in the inner regions of planetary systems. It is likely that the level of dust falls below our detection limits, before the number of particles in the inner regions falls to zero ($t_{zodi}$). Thus, we have probably overestimated the lifetime of the exozodi, $t_{zodi}$. In addition to this, if all the stars observed with exozodiacal dust were to be planetary systems in the aftermath of instabilities, we would also expect to see certain observable properties of the population, namely: 
\begin{itemize}
\item{{\it Significant variation in the levels of exozodiacal dust detected}

 The wide range of different instabilities and the steep decrease in the levels of scattered material directly following an instability, leads us to predict a wide range of different dust levels in planetary systems. The exozodi observed so far tend to be detected at the limit of our capabilities, so it is very hard to tell whether such a spread really exists. 

}
\item{{\it Potentially a variation in the levels of exozodiacal dust with time}

Following an instability we expect the levels of exozodiacal dust to fall off rapidly. Nonetheless, we anticipate that these timescales are still long compared to human timescales, being in some manner related to $t_{zodi}$ and that it is therefore unlikely that we are able to detect variation in the level of infrared excess on year (ten year) timescales. Variations such as those seen by \cite{Melis2012} are likely to have a different origin.

}

\item{{\it A decrease in the fraction of systems where exozodiacal dust is detected and the level of exozodiacal dust detected with the age of the system }

Instabilities have a higher probability to occur in planetary systems earlier in their evolution. No such age dependence was found in the observations of \cite{Absil2013}. 
}
\item{{\it A significant increase in the fraction of systems where exozodiacal dust is detected with stellar mass }

The shorter main-sequence lifetime of higher mass A stars increases the probability of catching a system in the aftermath of an instability. Although such a dependence is observed in \cite{Absil2013}, given the absence of the other predicted signatures, it seems more likely that the spectral type dependence has another origin, potentially in the more fundamental differences in the structure of planetary systems around stars of different spectral types and the different detection limits for exozodiacal dust. 

}
\item{{\it A correlation between the high levels of dust observed in the inner regions of the planetary system (exozodi) with high levels of dust throughout the planetary system.}

If a planetary system is undergoing an instability then material is scattered throughout the planetary system, not just into the inner regions. Thus, signatures of the instability should be observable in emission at longer wavelengths, as suggested by \cite{Booth09} and used to limit the incidence of LHB-like events to less than $12\%$ of sun-like stars. The results of \cite{Absil2013} could be interpreted to be in favour of such a scenario, given that 33\% of their sample with a cold, outer debris disc were found to have K-band excess, as opposed to 23\% of their sample for which there is no evidence for a cold, outer disc. Such a simple correlation provides no information regarding whether the dust is really spread throughout the planetary system, or in fact limited to a thin inner belt and a thin outer belt.

}
\end{itemize}

\section{Conclusions}
\label{sec:conclusion}
We have discussed a potential origin to the high levels of exozodiacal dust found in many planetary systems in dynamical instabilities, akin to the LHB. 
We have used N-body simulations to show that whilst material is almost always scattered into the inner regions of the planetary system following such an instability, this material has a short lifetime. We characterised the maximum possible lifetime of this material using the parameter $t_{zodi}$, finding a median value of $t_{zodi}=2.8$Myr ({\tt low mass}) and $t_{zodi}=0.7$Myr ({\tt high mass}), with a maximum value of $t_{zodi}=47$Myr. This suggests that an absolute maximum of $\le 1\%$ of sun-like stars, with 3-planets and planetesimals belts at 10-20AU, are likely to be observed in the aftermath of an instability. Given a median detection fraction of 0.06\% (0.01\%) for our {\tt low mass} ({\tt high mass}) simulations, we conclude that it is unlikely that we have observed a single sun-like star in the aftermath of a dynamical instability and we definitely cannot explain the $18\%$ of systems detected by \cite{Absil2013} at $2.2\mu$m and we struggle to even explain the 1\% detection rate in the mid-IR of \cite{Lawler2011}. Such a strong conclusion cannot be made for samples of very young stars or higher mass stars (A-stars), with their tendency to have planetary systems at larger orbital radii. However, we reiterate our conclusion that there is a very low probability of catching a planetary system in the observable aftermath of a dynamical instability.

\section{Acknowledgements} 

AB acknowledges the support of the ANR-2010 BLAN-0505-01 (EXOZODI). This work benefited from many useful discussions with Philippe Th\'{e}bault, Olivier Absil and the rest of the EXOZODI team. SNR thanks the CNRS's PNP program and the ERC for their support. We thank the referee for their help improving the manuscript.

\bibliographystyle{mn}

\bibliography{ref}

\end{document}